\documentclass[10pt]{iopart}
\usepackage{latexsym}
\usepackage{amsfonts}
\usepackage{amsbsy}
\usepackage{amssymb}   
\usepackage{accents}
\usepackage{hyperref}
\hypersetup{colorlinks=true,linkcolor=blue,urlcolor=blue,citecolor=blue}

\usepackage{tensor}
\usepackage{color}
\usepackage{cite}

\newcommand{\intprod}{\mbox{$\; \put(0,0){\line(1,0){.9}}
		\put(.9,0){\line(0,1){1.6}} \; \, \,  $}}

\begin{document}

\title[Reformulation of the symmetries of first-order general relativity]
{Reformulation of the symmetries of first-order general relativity}

\author{Merced Montesinos$^{1,2}$, Diego Gonz\'alez$^1$, Mariano Celada$^1$ and Bogar D\'{\i}az$^2$}

\address{$^1$ Departamento de F\'{\i}sica, Cinvestav, Avenida Instituto Polit\'ecnico Nacional 2508,
	San Pedro Zacatenco, 07360, Gustavo A. Madero, Ciudad de M\'exico, M\'exico.}
\address{$^2$ Departamento de Matem\'aticas, Instituto de Ciencias, Benem\'erita Universidad Aut\'onoma de Puebla, Ciudad 
	Universitaria, 72572, Puebla, Puebla, M\'exico.}

\eads{\mailto{merced@fis.cinvestav.mx},  \mailto{dgonzalez@fis.cinvestav.mx}, \mailto{mcelada@fis.cinvestav.mx} and  \mailto{bdiaz@alumnos.fcfm.buap.mx}
}


\begin{abstract}
	
We report a new internal gauge symmetry of the $n$-dimensional Palatini action with cosmological term ($n>3$) that is the generalization of three-dimensional local translations. This symmetry is obtained through the direct application of the converse of Noether's second theorem on the theory under consideration. We show that diffeomorphisms can be expressed as linear combinations of it and local Lorentz transformations with field-dependent parameters up to terms involving the variational derivatives of the action. As a result, the new internal symmetry together with local Lorentz transformations can be adopted as the fundamental gauge symmetries of general relativity. Although their  gauge algebra is open in general, it allows us to recover, without resorting to the equations of motion, the very well-known Lie algebra satisfied by translations and Lorentz transformations in three dimensions. We also report the analog of the new gauge symmetry for the Holst action with cosmological term, finding that it explicitly depends on the Immirzi parameter. The same result concerning its relation to diffeomorphisms and the open character of the gauge algebra also hold in this case. Finally, we consider the non-minimal coupling of a scalar field to gravity in $n$ dimensions and establish that the new gauge symmetry is affected by this matter field. Our results indicate that general relativity in dimension greater than three can be thought of as a gauge theory.
\end{abstract}


\section{Introduction}

Noether's theorems~\cite{Noether,EmmyNoether,Bessel-Hagen} are powerful and beautiful mathematical results linking continuous symmetries and conservations laws in theories with a variational principle. Noether's first theorem deals with systems whose symmetries depend on a finite (or infinite but countable) number of arbitrary parameters (such as global symmetries), whereas Noether's second theorem extends the results of the first to include infinite-dimensional symmetries (such as gauge symmetries), where the symmetry transformations are parametrized by arbitrary functions (and their derivatives). Although both theorems allow us to obtain on-shell conserved Noether currents from a given symmetry transformation of the action principle, they also work the other way around. In particular, the converse of Noether's second theorem can be used to uncover a gauge symmetry of the theory from a Noether identity. In light of this, in this paper we want to explore its implications on the first-order formulation of general relativity.

In the first-order formalism, general relativity is by construction invariant under local Lorentz transformations and diffeomorphisms. These two transformations actually comprise the set of gauge symmetries of the theory. Since diffeomorphisms move the points of the manifold, they are difficult to deal with at the quantum level, and so it would be desirable to supersede them by another symmetry that perhaps could be more manageable in that regime. According to the experience gained when dealing with internal gauge symmetries (such as the ones involved in Yang-Mills theory), a symmetry of this kind might be a better option to approach the problem of quantizing gravity, something hitherto not fully achieved. Notice that this program has worked in the three-dimensional setting, where giving up diffeomorphisms in favor of an internal gauge symmetry (local translations) has led to fruitful results~\cite{AchTow,Witten}. There have been several attempts to extend this strategy to higher dimensions~\cite{Utiyama,Kibble,Hehl,Cho,MacDowell} (see also~\cite{Blago1} and references therein), although the program has not been successfully completed. In particular, the transformation introduced in~\cite{Hehl} and rederived as a ``translational'' symmetry in the framework of the $n$-dimensional Einstein-Cartan action in~\cite{Kiriushcheva} does not even reproduce three-dimensional local translations off-shell.

In this paper we report the existence of a new internal gauge symmetry for general relativity in dimensions greater than three that is the genuine extension of three-dimensional local translations to higher dimensions. This symmetry results from the construction of a nontrivial Noether identity associated to the $n$-dimensional Palatini action with cosmological term, which, according to the converse of Noether's second theorem, encodes a gauge symmetry of the theory. Since diffeomorphisms and the new gauge symmetry are not independent, this symmetry together with local Lorentz transformations can be taken as a fundamental set to describe the full gauge symmetry of general relativity, diffeomorphisms then becoming a derived symmetry. The algebra generated by the new set of gauge symmetries turns out to be open~\cite{Henneaux,Henneaux199047}, {\it i.e.}, it closes with structure functions in the generic case and involves terms proportional to the variational derivatives. Although this sort of algebras is certainly complicated, they can be dealt with either by employing  the Batalin-Vilkovisky (BV) formalism~\cite{BATALIN198127,FUSTER-HENNEAUX2005} or by first closing the algebra~\cite{BATALIN1984106} and then using the standard BRST approach~\cite{Henneaux},  which could provide new ideas to quantize gravity. The structure of the new internal gauge symmetry reported in this paper depends on the spacetime dimension and off-shell is different from that of references~\cite{Hehl,Kiriushcheva}. Furthermore, in contrast to diffeomorphisms and local Lorentz transformations, which always take the same form disregarding the action principle, the new gauge symmetry is sensitive to the structure of the Lagrangian underlying general relativity. This will be exhibited in four dimensions in the second part of this paper, where, in addition to the Palatini action, we also have the Holst action~\cite{Holst}. There, we shall see that the new gauge symmetry explicitly depends on the Immirzi parameter for the latter. Moreover, we also show that the new gauge symmetry is sensitive to the coupling of matter fields to general relativity. This is explicitly displayed for the non-minimal coupling of a scalar field $\phi$ to gravity in $n$ dimensions.

\section{Internal gauge symmetries of the $n$-dimensional Palatini action.}

Let us consider an orientable $n$-dimensional manifold $\mathcal{M}^n$ (with $n\geq 3$) . In order to be more general, we address both Lorentzian and Euclidean manifolds at once, and hence we denote the internal group by $SO(\sigma)$, where $SO(-1)=SO(1,n-1)$ for Lorentzian manifolds ($\sigma=-1$) and $SO(+1)=SO(n)$ for Euclidean ones ($\sigma=+1$). The standard picture of general relativity in $n$ dimensions is provided by the Palatini or Einstein-Cartan action with cosmological term $S[e,\omega]=\int_{\mathcal{M}^n} L$ , whose Lagrangian $n$-form $L$ is
\begin{eqnarray}\label{palatinindim}
L = \kappa \left [ \star (e_I \wedge e_J) \wedge R^{IJ} [\omega] - \frac{2 \Lambda}{n!} \epsilon_{I_1\dots I_n} e^{I_1} \wedge \!\dots\!\wedge e^{I_n} \right ],
\end{eqnarray}
where $e^I$ is an orthonormal frame of 1-forms, $\omega^I{}_J$ is an $SO(\sigma)$ connection 1-form with curvature $R^I{}_J[\omega]:= d \omega^I{}_J + \omega^I{}_K \wedge \omega^K{}_J$, $\kappa$ is a constant (whose dimensions depend on $n$), $\Lambda$ is the cosmological constant,  and $\star$ is the Hodge dual operator:
\begin{eqnarray}
\star(e_{I_1} \wedge \dots\wedge e_{I_k}) =\frac{1}{(n-k)!} \epsilon_{I_1\dots I_kI_{k+1} \dots I_n} e^{I_{k+1}} \!\wedge \dots \wedge e^{I_n}. \nonumber
\end{eqnarray}
The internal indices $I,J,\dots$ take the values $0,1,\dots,n-1$ and are raised and lowered with the internal metric $(\eta_{IJ})=\rm{diag}(\sigma,1,\dots,1)$. Moreover, the internal tensor $\epsilon_{I_1\dots I_n}$ is totally antisymmetric and satisfies $\epsilon_{01\dots n-1}=1$. Our convention for the antisymmetrizer is $A^{[IJ]}:=(A^{IJ}-A^{JI})/2$. The variational derivatives of the action defined by (\ref{palatinindim}) with respect to the independent variables are
\numparts
\begin{eqnarray} 
	\fl &&{\mathcal E}_I :=\frac{\delta S}{\delta e^I}= (-1)^{n-1} \kappa \star 
	(e_I\wedge e_J\wedge e_K) 
	\wedge \left [ R^{JK} [\omega] - \frac{2\Lambda}{(n-1)(n-2)} e^J \wedge e^K \right ], \label{moteqn1}\\
	\fl &&{\mathcal E}_{IJ}:= \frac{\delta S}{\delta \omega^{IJ}}=(-1)^{n-1} \kappa \, D \star (e_I \wedge e_J),\label{moteqn2}
\end{eqnarray}
\endnumparts
with $D$ being the $SO(\sigma)$ covariant derivative. The equations of motion correspond to ${\mathcal E}_I=0$ and ${\mathcal E}_{IJ}=0$, which lead to Einstein's equations with cosmological constant. Nevertheless, since our approach is off-shell, ${\mathcal E}_I$ and ${\mathcal E}_{IJ}$ are nonvanishing in general.

As is clear from the structure of (\ref{palatinindim}), the Lagrangian is a Lorentz scalar, and hence the action is invariant under local Lorentz transformations. Furthermore, under an active diffeomorphism $\Psi: \mathcal{M}^n\rightarrow \mathcal{M}^n$ the frame and the connection are pulled back by the induced map $\Psi^{\ast}$, and then the Lagrangian fulfills $L[\Psi^{\ast} e^I, \Psi^{\ast} \omega ] = \Psi^{\ast} ( L [ e, \omega] )$. Therefore, by Stokes theorem, the action is also diffeomorphism-invariant. These two transformations comprise the set of fundamental symmetries of general relativity. Infinitesimally (or for transformations near the identity), local Lorentz transformations and diffeomorphisms of $e^I$ and $\omega^{IJ}$ are:
\numparts
\begin{eqnarray}
\rm{Lorentz}:&\quad
\delta_{\tau}e^I =\tau^I{}_{J} e^J ,\quad &\delta_{\tau}\omega^{IJ}=- D \tau^{IJ},\label{lorenztt}\\
\rm{Diffeos}:&\quad\delta_{\xi} e^I = {\mathcal L}_{\xi} e^I,\quad &\delta_{\xi} \omega^{IJ}= {\mathcal L}_{\xi} \omega^{IJ}, \label{diffeos}
\end{eqnarray}
\endnumparts
where $\tau^{IJ}(=-\tau^{JI})$ is an $\mathfrak{so}(\sigma)$-valued function (the gauge parameter) and ${\mathcal L}_{\xi}$ is the Lie derivative along the generator $\xi$ of the diffeomorphism.

In general, an infinitesimal transformation of the fields depending on arbitrary functions (and their derivatives) is said to be a gauge symmetry of the action if the Lagrangian is quasi-invariant under it, {\it i.e.}, if the Lagrangian remains invariant up to a total derivative~\cite{Noether,EmmyNoether,Bessel-Hagen} (see also~\cite{Henneaux,Henneaux199047}). This is true, for instance, for the previous infinitesimal transformations, in whose case the change of the Lagrangian (\ref{palatinindim}) vanishes for local Lorentz transformations and is equal to the differential of an $(n-1)-$form for diffeomorphisms. Bearing this definition in mind, we have the following statement:

\textit{\textbf{Theorem 1}}. Let $\rho^I(x)$ be arbitrary gauge parameters, where $x$ are local coordinates on $\mathcal{M}^n$. Then, the infinitesimal internal transformation 
\begin{eqnarray}\label{gaugetrn}
\delta_{\rho} e^I &&= d\rho^I+\omega^I{}_J\rho^J\hspace{2mm} (\equiv D \rho^I), \nonumber\\
\delta_{\rho} \omega^{IJ} &&= \frac{\sigma (n-3)}{(n-2)!} \Bigl( \epsilon^{IJLK_1 \dots K_{n-3}}\ast \mathcal{R}_{M K_1 \dots K_{n-3}LN} \nonumber\\
&& + \ast\mathcal{R}\ast _{K_1 \dots K_{n-4}MN}{}^{K_1 \dots K_{n-4}IJ} \Bigr) \rho^M e^N+ \frac{2 \Lambda}{n\!-\!2} \rho^{[I} e^{J]},
\end{eqnarray}
is a gauge symmetry of the the $n$-dimensional Palatini action with cosmological term. Here, we have expressed the curvature as $R^{IJ}[\omega]:=(1/2)\mathcal{R}^{IJ}{}_{KL} e^K\wedge e^L$, while
\numparts
\begin{eqnarray}
\ast\mathcal{R}_{I_1 \dots I_{n-2}MN}  &:=& \frac12 \epsilon_{I_1 \dots I_{n-2}KL} \mathcal{R}^{KL}{}_{MN}, \\
\mathcal{R}\ast{}^{MN I_1 \dots I_{n-2}} &:=& \frac12 \epsilon^{I_1 \dots I_{n-2}KL} \mathcal{R}^{MN}{}_{KL}
\end{eqnarray}
\endnumparts
define the left and right internal duals, respectively. 

\textit{\textbf{Proof}}. The change of (\ref{palatinindim}) under the transformation (\ref{gaugetrn}) reads

\begin{eqnarray}
\fl \delta_{\rho} L = d\left\{\frac{\kappa}{n-2} \rho^I \star 
(e_I\wedge e_J\wedge e_K)\wedge\left[ R^{JK} [\omega]  + \frac{2 \Lambda}{(n-1)(n-2)} e^J \wedge e^K \right]\right\}.\label{changeL}
\end{eqnarray}
Therefore, the Lagrangian changes by a total derivative and, as result, the transformation (\ref{gaugetrn}) {\it is} a gauge symmetry of the action defined by (\ref{palatinindim}). \hfill $\blacksquare$

\textit{\textbf{Converse of Theorem 1}}. By computing the covariant derivative of (\ref{moteqn1}), the following Noether identity emerges:
\begin{eqnarray}\label{generalidentityn1}
D {\mathcal E}_I -Z_n{}^{KL}{}_{IJ}e^J \wedge \mathcal{E}_{KL}  =0, 
\end{eqnarray}
with
\begin{eqnarray}
Z_n{}^{IJ}{}_{KL}:=&&\frac{\sigma (n-3)}{(n-2)!} \Bigl( \epsilon^{IJNM_1 \dots M_{n-3}} \ast\!\mathcal{R}_{K M_1 \dots M_{n-3}NL} \nonumber\\
&&+ \ast\mathcal{R}\!\ast _{M_1 \dots M_{n-4}KL}{}^{M_1 \dots M_{n-4}IJ} \Bigr) + \frac{2 \Lambda}{n\!-\!2} \delta^{[I}_K \delta^{J]}_L.\label{Zn}
\end{eqnarray}
Multiplying (\ref{generalidentityn1}) by the parameter $\rho^I$, we arrive at the off-shell identity 
\begin{eqnarray}\label{generalidentityn}
	{\mathcal E}_I \wedge \underbrace{D\rho^I }_{\delta_{\rho} e^I}+ {\mathcal E}_{IJ} \wedge \underbrace{Z_n{}^{IJ}{}_{KL}\rho^K e^L}_{\delta_{\rho} \omega^{IJ}} + d \left [ (-1)^n \rho^I {\mathcal E}_I\right ]=0,
\end{eqnarray}
where the $(n-1)$-form inside the square brackets is recognized as the Noether current \cite{Noether,EmmyNoether,Bessel-Hagen}. According to the converse of Noether's second theorem, the expressions accompanying the variational derivatives in (\ref{generalidentityn}) allow us to read off the gauge symmetry associated to (\ref{generalidentityn1}), which is exactly that given by  (\ref{gaugetrn}). \hfill  $\blacksquare$

We point out that an analogous procedure can be used to uncover the local Lorentz symmetry of the theory. Indeed, if we first take the covariant derivative of (\ref{moteqn2}), we obtain the Noether identity
\begin{eqnarray}\label{loretzgaugeiden}
D {\mathcal E}_{IJ} - e_{[I} \wedge {\mathcal E}_{J]}=0,
\end{eqnarray}
which, after multiplied by the gauge parameter $\tau^{IJ}$, leads to the transformation (\ref{lorenztt}). Thus, (\ref{loretzgaugeiden}) is the Noether identity associated to local Lorentz symmetry.

One important fact about the new internal gauge symmetry is that it is not independent of spacetime diffeomorphisms. This is to be expected, since diffeomorphisms and local Lorentz transformations comprise a set of fundamental gauge symmetries of general relativity, and so any other gauge symmetry of the theory must be expressible in terms of them. To make this precise, let us note that  $Z_n{}^{IJ}{}_{KL}$ can be alternatively written as
\begin{eqnarray}\label{zetan}
\fl Z_n{}^{IJ}{}_{KL} = {\mathcal R}^{IJ}{}_{KL} - 2\delta^{[I}_K {\mathcal R}^{J]}{}_L +\frac{2}{n-2} \delta^{[I}_L {\mathcal R}^{J]}{}_K + \frac{1}{n-2} \left ( {\mathcal R} + 2 \Lambda \right ) \delta^{[I}_K \delta^{J]}_L,
\end{eqnarray}
with ${\mathcal R}^{I}{}_J:={\mathcal R}^{KI}{}_{KJ}$ and ${\mathcal R}:={\mathcal R}^{I}{}_I$ the Ricci tensor\footnote{Here, the Ricci tensor is in general nonsymmetric because we are working off-shell. It is symmetric only on-shell.} and the scalar curvature, respectively. Using this expression together with Cartan's formula, it can be shown that infinitesimal diffeomorphisms are linear combinations of both local Lorentz transformation and the new gauge transformation (\ref{gaugetrn}) with respective field-dependent gauge parameters $\tau^{IJ}= - \xi \, \intprod \omega^{IJ}$ and $\rho^I = \xi \, \intprod e^I$ (``$\intprod$'' stands for contraction~\cite{Gerardo}) up to terms proportional to the variational derivatives:
\numparts
\begin{eqnarray}
& &\delta_{\xi} e^I = (\delta_{\tau}+\delta_{\rho})e^I+\mbox{terms proportional to ${\mathcal E}_{IJ}$}, \label{diffesym1}\\
& &\delta_{\xi} \omega^{IJ} = (\delta_{\tau}+\delta_{\rho})\omega^{IJ} +\mbox{terms proportional to ${\mathcal E}_{I}$}.\label{diffesym2}
\end{eqnarray}
\endnumparts
Therefore, diffeomorphisms (near the identity) can be constructed out of these two symmetries. This implies that we can adopt, instead of local Lorentz symmetry and diffeomorphisms, the set composed of local Lorentz symmetry and the new gauge transformation (\ref{gaugetrn}) to describe the full gauge invariance of general relativity {\it without} compromising its propagating degrees of freedom. Consequently, in the first-order formalism, general relativity in any dimension can be thought of as a gauge theory whose internal symmetry is given by (\ref{lorenztt}) along with (\ref{gaugetrn}). In this framework, diffeomorphisms are nothing but a {\it derived} symmetry.

Now we must check the algebra of gauge transformations. By computing the commutators among $\delta_{\tau}$ and $\delta_{\rho}$ acting on both the frame and the connection, the algebra reads
\numparts
\begin{eqnarray}
\left [ \delta_{\tau_1}, \delta_{\tau_2} \right ] &=& \delta_{\tau_3} \quad (\tau_3^{IJ}:=2\tau_1^{[I|K}  \tau_2^{|J]}{}_K), \label{algebrapalatnd1}\\
\left [ \delta_{\rho}, \delta_{\tau} \right ] &=& \delta_{\rho_1} \quad (\rho_1^I:=\tau^I{}_J\rho^J), \label{algebrapalatnd2}\\
\left [ \delta_{\rho_1}, \delta_{\rho_2} \right ] &=& \delta_{\tau} + \mbox{terms involving ${\mathcal E}_I$ and ${\mathcal E}_{IJ}$}, \label{algebrapalatnd3}
\end{eqnarray}
\endnumparts
where $\tau^{IJ} := 2 Z_n{}_K{}^{[I}{}_L{}^{J]} \rho^{[K}_1 \rho^{L]}_2$ in the last line. Equation (\ref{algebrapalatnd1}) manifests the fact that (local) Lorentz transformations by themselves form a group. The relation (\ref{algebrapalatnd2}) indicates that the commutator of the new gauge transformation (\ref{gaugetrn}) and a Lorentz transformation is a transformation of the same type as (\ref{gaugetrn}) whose gauge parameter has been rotated by the Lorentz transformation. Finally, (\ref{algebrapalatnd3}) tells us that up to terms proportional to the variational derivatives, the commutator of two gauge transformations (\ref{gaugetrn}) is a Lorentz transformation with a field-dependent gauge parameter as was identified there. For $n>3$, the commutator algebra among $\delta_{\tau}$ and $\delta_{\rho}$ closes with structure functions, but since the last commutator involves terms proportional to the variational derivatives (which generate trivial transformations), the algebra of gauge transformations is {\it open}~\cite{Henneaux}. The case $n=3$ will be discussed in the following two paragraphs.

\subsection*{The new gauge symmetry for $n=3$.}

Let us see in detail the form that the new gauge symmetry takes in $n=3$ spacetime dimensions. In this case, the Lagrangian (\ref{palatinindim}) takes the simpler form
\begin{eqnarray}\label{palatini3d1}
L = \kappa \, \epsilon_{IJK} e^I \wedge \left( R^{JK}[\omega] - \frac{\Lambda}{3} e^J \wedge e^K\right).
\end{eqnarray}
Setting $n=3$ in (\ref{gaugetrn}), we see that the terms involving the components of the Riemann tensor drop out. The resulting transformation reads
\begin{eqnarray}\label{gaugetr3}
\delta_{\rho} e^I = D \rho^I, \qquad \delta_{\rho} \omega^{IJ} =2\Lambda\rho^{[I} e^{J]}.
\end{eqnarray}
In the context of three-dimensional gravity this transformation is very well-known. There, it is referred to as ``local translations''~\cite{Carlip2+1} and has associated the frame as the gauge field. Accordingly, the gauge transformation (\ref{gaugetrn}) \textit{corresponds to the higher-dimensional generalization} of three-dimensional local translations. Notice that the change of (\ref{palatini3d1}) under the transformation (\ref{gaugetr3}) yields
\begin{eqnarray}
\delta_{\rho} L = d\left\{ \kappa \rho^I \epsilon_{IJK} \left( R^{JK} [\omega]  + \Lambda e^J \wedge e^K \right)\right\},\label{changeL3D}
\end{eqnarray}
which can be obtained from (\ref{changeL}) by setting $n=3$. Furthermore, in this case (\ref{Zn}) collapses to $Z_3{}^{IJ}{}_{KL}:=2 \Lambda\delta^{[I}_K \delta^{J]}_L$ and then the Noether identity (\ref{generalidentityn1}) corresponding to (\ref{gaugetr3}) acquires the form
\begin{eqnarray}\label{generalidentityn13D}
D {\mathcal E}_I -2 \Lambda e^J \wedge \mathcal{E}_{IJ}  =0.
\end{eqnarray}

On the other side, by fixing $n=3$ in the gauge algebra (\ref{algebrapalatnd1})-(\ref{algebrapalatnd3}), we observe that (\ref{algebrapalatnd1}) and (\ref{algebrapalatnd2}) remain unchanged, while (\ref{algebrapalatnd3}) reduces to 
\begin{eqnarray}\label{algebra3d}
\left [ \delta_{\rho_1}, \delta_{\rho_2} \right ] &=& \delta_{\tau} \quad ( \tau^{IJ} := 2 \Lambda \rho^{[I}_1 \rho^{J]}_2 ),
\end{eqnarray}
this because for $n=3$ the terms proportional to the variational derivatives in (\ref{algebrapalatnd3}) are absent. Therefore, for $n=3$ the commutators (\ref{algebrapalatnd1}), (\ref{algebrapalatnd2}), and (\ref{algebra3d}) define a true Lie algebra that can be recognized, for $\sigma=-1$, as the Lie algebra of the de Sitter group $SO(1,3)$ if $\Lambda > 0$, the anti-de Sitter group $SO(2,2)$ if $\Lambda<0$, and the Poincar\'e group $ISO(1,2)$ if $\Lambda=0$. In this setting, the frame and the three-dimensional Lorentz connection can be combined into an enlarged connection implementing the gauge invariance of the theory under one of the previous Lie groups. As a result, the integrand in the action (\ref{palatini3d1}) can be rewritten as a Chern-Simons form for the enlarged connection~\cite{AchTow,Witten}, which is an archetypal gauge theory. Consequently, three-dimensional general relativity is a gauge theory of gravity (it can also be interpreted as a topological $BF$ theory; see \cite{cqgrevBF}). It is worth pointing out that the relations (\ref{diffesym1})-(\ref{diffesym2}) also hold in three dimensions (see for instance~\cite{Carlip2+1}), being the exchange of diffeomorphisms for an internal gauge symmetry what allows three-dimensional general relativity to be expressed as a true gauge theory.
  
\subsection*{Comparison with another set of internal transformations.}

Attempts to generalize three-dimensional local translations to higher dimensions have given rise to the so-called Poincar\'e gauge theory, where several candidates of transformations have been proposed~\cite{Utiyama,Kibble,Hehl,Cho}. In what follows we focus our attention on the symmetry of~\cite{Hehl}, which was generalized to the $n$-dimensional Einstein-Cartan action in~\cite{Kiriushcheva}. Notice, however, that those works only addressed the case $\Lambda=0$, whereas in our approach we also encompass the case of nonvanishing cosmological constant. It turns out that the same ``translational'' transformation of references~\cite{Hehl,Kiriushcheva} also works for $\Lambda\neq0$, which in our notation can be written as
\begin{eqnarray}
\delta_\rho e^I= D\rho^I + \rho \intprod De^I, \qquad \delta_\rho \omega^{IJ}={\mathcal R}^{IJ}{}_{KL} \rho^K e^L, \label{HehlTrans}
\end{eqnarray}
where $\rho:=\rho^I \partial_I$ and $\partial_I$ is the dual basis of $e^I$ ($\partial_I\intprod e^J=\delta_I^J$). This symmetry together with local Lorentz transformations could also be taken as a set of fundamental transformations to describe the full gauge symmetry of general relativity. 

Let us deduce (\ref{HehlTrans}). Using ({\ref{moteqn1}) and (\ref{moteqn2}) we get the Noether identity
\begin{eqnarray}
D {\mathcal E}_I - (\partial_I \intprod De^J)\wedge {\mathcal E}_J - (\partial_I \intprod R^{JK})\wedge {\mathcal E}_{JK} =0. \label{NoetherHehl}
\end{eqnarray}
Multiplying this relation by the arbitrary parameter $\rho^I$ we obtain the off-shell identity
\begin{eqnarray}
{\mathcal E}_I \wedge \delta_\rho e^I + {\mathcal E}_{IJ}  \wedge \delta_\rho \omega^{IJ} + d \left [ (-1)^n \rho^I {\mathcal E}_I \right ] =0,
\end{eqnarray}
with $\delta_\rho e^I$ and $\delta_\rho \omega^{IJ}$ those given in (\ref{HehlTrans}).

Some remarks are in order:
\begin{enumerate}
	\item As can be seen, the transformation (\ref{HehlTrans}) and the new internal gauge transformation (\ref{gaugetrn}) have some structural differences off-shell. With respect to the transformation of the frame, we see that (\ref{HehlTrans}) possesses an extra term proportional to $De^I$ that is absent in (\ref{gaugetrn}). Regarding the transformation of the connection, the transformation (\ref{gaugetrn}) includes some curvature terms together with a term proportional to the cosmological constant that are not present in (\ref{HehlTrans}). In addition, the spacetime dimension $n$ does not enter in (\ref{HehlTrans}), whereas it explicitly appears in (\ref{gaugetrn}).
	\item Because of the structural form of  (\ref{HehlTrans}), when $n=3$ this transformation {\it does not}  yield, off-shell, three-dimensional local translations [see (\ref{gaugetr3})]. In contrast, the transformation (\ref{gaugetrn}) actually does it, as was shown above. This agrees with the fact that, according to Noether's theorem, gauge symmetries must be treated off-shell~\cite{Noether,EmmyNoether}.
	\item  As a consequence of item (ii), the transformation (\ref{HehlTrans}) {\it is not} the generalization to higher dimensions of the transformation (\ref{gaugetr3}). The generalization of (\ref{gaugetr3}) is indeed given by the new internal gauge transformation (\ref{gaugetrn}), which is its main strength.
\end{enumerate}

Now, we extend our approach to Holst action in the following section.

\section{Internal gauge symmetries of the Holst action.}

One interesting feature of the new internal gauge symmetry is that, off-shell, its form depends on the Lagrangian underlying general relativity or, more precisely, on its corresponding variational derivatives. It is well-known that four-dimensional general relativity in the first-order formalism can be equivalently described by the Holst Lagrangian~\cite{Holst}, which plays a fundamental role in the loop approach to quantum gravity~\cite{RoveLew,Rovebook,Thiebook,perez2013}. By including a cosmological term, the Lagrangian is
\begin{eqnarray}\label{holst1}
L = \kappa e^I\wedge e^J \wedge \left ( P_{IJKL} R^{KL}  - \frac{\Lambda}{12}  \epsilon_{IJKL}e^K\wedge e^L\right ),
\end{eqnarray}
where $P_{IJKL}:= (1/2) \epsilon_{IJKL} + (\sigma/\gamma) \eta_{[I|K} \eta_{|J]L}$ and $\gamma\in\mathbb{R}-\{0\}$ is the Immirzi parameter~\cite{Barbero, Immirzi}. This Lagrangian is obtained from (\ref{palatinindim}) (with $n=4$) by adding the Holst term $e^I\wedge e^J\wedge R_{IJ}$, which defines a topological field theory~\cite{Liu}. The analogs of (\ref{moteqn1}) and (\ref{moteqn2}) read
\numparts
\begin{eqnarray}
{\mathcal E}_I &:=& - 2 \kappa e^J \wedge \left( P_{IJKL} R^{KL} - \frac{\Lambda}{6} \epsilon_{IJKL}e^K\wedge e^L\right),\label{eqmoth1} \\
{\mathcal E}_{IJ} &:=& - \kappa D (P_{IJKL} e^K \wedge e^L). \label{eqmoth2}
\end{eqnarray}
\endnumparts
On-shell, these quantities lead to Einstein's equations, and that is the reason why (\ref{holst1}) also describes general relativity.

The action defined by (\ref{holst1}) is invariant under both local Lorentz transformations and spacetime diffeomorphisms. Furthermore, we have the following assertion:

\textit{\textbf{Theorem 2}}. The infinitesimal internal transformation
\begin{eqnarray}
	\delta_{\rho}e^I = && D \rho^I, \nonumber\\
	\delta_{\rho} \omega^{IJ} = && (P^{-1})^{IJPQ} \left [ \frac12 P_{KLMN} \mathcal{R}^{MN}{}_{\!\!PQ}\right.\nonumber\\
	&& \left.- P_{KPMN} \mathcal{R}^{MN}{}_{\!\!\!QL}  + \frac{\Lambda}{3}(P_{KLPQ} + \!2 P_{KPQL}) \right ] \rho^K e^L, \label{tresholstsym} 
\end{eqnarray}
and $(P^{-1})^{IJKL}$ satisfying $(P^{-1})^{IJKL}P_{KLMN}=\delta^I_{[M}\delta^J_{N]}$, is a gauge symmetry of the action defined by (\ref{holst1}).

\textit{\textbf{Proof}}. Using (\ref{tresholstsym}), the change of (\ref{holst1}) is 
\begin{eqnarray}
\delta_{\rho} L= d \left[ \kappa \rho^I e^J \wedge \left(P_{IJKL} R^{KL} + \frac{\Lambda}{6} \epsilon_{IJKL}e^K\wedge e^L\right) \right],
\end{eqnarray}
which shows that it is quasi-invariant under (\ref{tresholstsym}). Therefore, this transformation {\it is} a gauge symmetry of the Holst action with cosmological constant. \hfill $\blacksquare$

\textit{\textbf{Converse of Theorem 2}}. By taking the covariant derivative of (\ref{eqmoth1}), we arrive at the Noether identity
\begin{eqnarray}
	D {\mathcal E}_I -Z^{KL}{}_{IJ}e^J \wedge \mathcal{E}_{KL}  =0, \label{gaugeholst}
\end{eqnarray}
where 
\begin{eqnarray}\label{Zholst}
\fl Z^{IJ}{}_{\!KL} \!=\!	(P^{-1})^{IJPQ} \! \left [ \frac12 P_{KLMN} \mathcal{R}^{MN}{}_{\!\!PQ}- P_{KPMN} \mathcal{R}^{MN}{}_{\!\!QL}  + \frac{\Lambda}{3}(P_{KLPQ} + \!2 P_{KPQL}) \right ]\!,\nonumber\\
\end{eqnarray}
which, after multiplied by the parameter $\rho^I$, leads precisely to the identity 
\begin{eqnarray}\label{holstgauge}
	{\mathcal E}_I \wedge \underbrace{D\rho^I }_{\delta_{\rho} e^I}+ {\mathcal E}_{IJ} \wedge \underbrace{Z{}^{IJ}{}_{KL}\rho^K e^L}_{\delta_{\rho} \omega^{IJ}} + d \left [ \rho^I {\mathcal E}_I\right ]=0,
\end{eqnarray}
where ${\mathcal E}_I$ and ${\mathcal E}_{IJ}$ are given by (\ref{eqmoth1}) and (\ref{eqmoth2}). Appealing to the converse of Noether's second theorem again, the terms next to the  previous variational derivatives correspond to the gauge transformation generated by (\ref{gaugeholst}), which matches (\ref{tresholstsym}). \hfill $\blacksquare$

As in the case of (\ref{gaugetrn}), diffeomorphisms can also be expressed as a linear combination of a local Lorentz transformation and the transformation (\ref{tresholstsym}) with field-dependent parameters [the same as in (\ref{diffesym1}) and (\ref{diffesym2})] modulo terms involving the variational derivatives (\ref{eqmoth1}) and (\ref{eqmoth2}). Therefore, we can take the latter together with local Lorentz invariance as the fundamental symmetries of the action determined by (\ref{holst1}) (and hence of general relativity). The commutator algebra among them is then the same as (\ref{algebrapalatnd1}), (\ref{algebrapalatnd2}), and (\ref{algebrapalatnd3}) with $Z^{IJ}{}_{KL}$ instead of $Z_4{}^{IJ}{}_{KL}$ and the variational derivatives those given in (\ref{eqmoth1}) and (\ref{eqmoth2}). Thereby, the algebra of gauge symmetries is also open. 

Note that the transformation of the frame in (\ref{tresholstsym}) is the same as in (\ref{gaugetrn}). Further, we can rewrite (\ref{Zholst}) as 
\begin{eqnarray}\label{tresholstsym1}
Z^{IJ}{}_{KL}
= \mathcal{R}^{IJ}{}_{KL} + (P^{-1})^{IJMN} \left(\ast X_{MNKL}+ \frac{\sigma}{\gamma}  Y_{MNKL}\right), 
\end{eqnarray}
where we have defined
\begin{eqnarray}
X_{IJKL}:=-2 \eta_{[I|K}{\mathcal R}_{|J]L} + \eta_{[I|L}{\mathcal R}_{|J]K} +\frac12 ( {\mathcal R} + 2 \Lambda ) \eta_{[I|K} \eta_{|J]L},\nonumber\\
Y_{IJKL}:=\frac12 \left( B_{JLIK} + B_{LIKJ} + B_{IKJL} \right),\label{defX1}
\end{eqnarray}
for $ R^I{}_J \wedge e^J =: (1/3!) B^I{}_{JKL} e^J \wedge e^K \wedge e^L$. As can be seen from (\ref{tresholstsym1}), the new gauge symmetry for the Holst action now explicitly depends on the Immirzi parameter, which means that, off-shell, this transformation is different from the one associated to the four-dimensional Palatini action [compare (\ref{tresholstsym1}) with {(\ref{zetan})}], even though at the Lagrangian level both actions describe the same classical physics\footnote{Nevertheless, the Hamiltonian form of the Holst action explicitly contains the Immirzi parameter, which later becomes significant at the quantum level since both the spectra of geometric operators~\cite{RoveSmolinNucP442,AshtLewacqg14,RovThiePRD57} and the black hole entropy~\cite{RovPRLt.77.3288,AshBaezcqg105,Meisscqg21,Agulloetcprl100,EngleNuiPRL105} depend on it.}. This is a consequence of the fact that the variational derivatives differ in each case, since they catch the structure of the Lagrangian underlying the theory. Finally, for $\gamma\rightarrow\infty$ we have $Z^{IJ}{}_{KL}|_{\gamma\rightarrow\infty}=Z_4{}^{IJ}{}_{KL}$, something expected since in that limit the Holst action collapses to Palatini's.

\section{Internal gauge symmetries of a non-minimally coupled scalar field}

Let us now study the case where a matter field is coupled to general relativity in the first-order formalism, which will illustrate how its presence affects the form of the new gauge symmetry. Here we concentrate on the non-minimal coupling of a scalar field $\phi$ to gravity in $n$ dimensions. The Lagrangian of the theory is given by
\begin{eqnarray}\label{GRmnsf}
\fl L_{\rm{SF}} = &\kappa \left\{ f(\phi)\star (e_I \wedge e_J) \wedge R^{IJ} [\omega] - 2 \Lambda \eta \right\}+\alpha\left[\frac12 K(\phi)d\phi\wedge\star d\phi-V(\phi)\eta\right],
\end{eqnarray}
where $\eta:=(1/n!)\epsilon_{I_1\dots I_n} e^{I_1} \wedge \!\dots\!\wedge e^{I_n}$ is the volume form. Likewise, $f$ (we assume $f>0$), $K$ and $V$ are arbitrary functions depending only on the scalar field, while $\alpha$ is a real parameter (that can be set equal to one without loss of generality). Notice that the cosmological term can be eliminated through a redefinition of the scalar field potential, but we shall keep this form since the first term on the right hand side of (\ref{GRmnsf}) resembles (\ref{palatinindim}). We point out that when passing to the second-order formalism, the curvature term of (\ref{GRmnsf}) produces a term proportional to the kinetic term $d\phi\wedge\star d\phi$, and so the arbitrariness in $K$ can be used to get rid of this contribution~\cite{Ashcalar}. The variational derivatives of the action defined by (\ref{GRmnsf}) then read
\numparts
\begin{eqnarray} 
\fl &&{\mathcal E}_I :=\frac{\delta S_{\rm{SF}}}{\delta e^I}= (-1)^{n-1} \kappa \star 
(e_I\wedge e_J\wedge e_K) 
\wedge \left [ f\  R^{JK} [\omega] - \frac{2\Lambda}{(n-1)(n-2)} e^J \wedge e^K \right ]\nonumber\\
\fl&&\hspace{21mm}+ (-1)^{n-1} T_{JI}\star e^J,\label{moteqnsf1}\\
\fl &&{\mathcal E}_{IJ}:= \frac{\delta S_{\rm{SF}}}{\delta \omega^{IJ}}=(-1)^{n-1} \kappa \, D [f \star (e_I \wedge e_J)],\label{moteqnsf2}\\
\fl &&{\mathcal E}_{\phi} :=\frac{\delta S_{\rm{SF}}}{\delta \phi}= -\alpha d(K\star d\phi)-\alpha \frac{dV}{d\phi}\eta+\kappa \frac{df}{d\phi}\star (e_I \wedge e_J) \wedge R^{IJ}+\frac{\alpha}{2}dK\wedge\star d\phi,\label{moteqnsf3}
\end{eqnarray}
\endnumparts
where
\begin{eqnarray}\label{EMtensor}
	T_{IJ}:=\alpha\left[-K\partial_I\phi\partial_J\phi+\left(\frac12 K \partial_L\phi\partial^L\phi-V\right)\eta_{IJ}\right],
\end{eqnarray}
for $\partial_I\phi:=\partial_I\intprod d\phi$, is the energy-momentum tensor of the scalar field, which is symmetric. Because of the presence of $f$ inside the covariant derivative  of (\ref{moteqnsf2}), the connection $\omega^{IJ}$ is no longer torsion-free on-shell (unless $f=\rm{constant}$, which constitutes, for $K=\rm{constant}$, the minimal coupling of the scalar field to general relativity).

The action defined by (\ref{GRmnsf}) is invariant under both local Lorentz transformations and diffeomorphisms. However, it also has an internal gauge symmetry of the same type described in the two previous sections that is related to diffeomorphisms in a similar fashion. Surprisingly, this gauge symmetry hinges on the way the scalar field couples to gravity. Indeed,  under the gauge transformation
\begin{eqnarray}\label{ngtsf}
	& \delta_{\rho} e^I=D\rho^I+\frac{1}{n-2}\rho^J\partial_J\intprod [e^I\wedge d(\ln f)],\nonumber\\
	& \delta_{\rho}\omega^{IJ}=(G_n{}^{IJ}{}_{KL}+F_n{}^{IJ}{}_{KL})\rho^K e^L,\nonumber\\
	& \delta_{\rho}\phi=\rho^I\partial_I \phi,
\end{eqnarray}
with
\begin{eqnarray}\label{Zsf}
	\fl G_n{}^{IJ}{}_{KL} = {\mathcal R}^{IJ}{}_{KL} - 2\delta^{[I}_K {\mathcal R}^{J]}{}_L +\frac{2}{n-2} \delta^{[I}_L {\mathcal R}^{J]}{}_K + \frac{1}{n-2} \left ( {\mathcal R} + 2 \Lambda f^{-1}\right ) \delta^{[I}_K \delta^{J]}_L,\label{Gsf}\\
	\fl F_n{}^{IJ}{}_{KL} = (\kappa f)^{-1}\left(\delta^{[I}_K T^{J]}{}_L -\frac{1}{n-2} \delta^{[I}_L T^{J]}{}_K - \frac{1}{n-2} T^{M}{}_M\delta^{[I}_K \delta^{J]}_L\right),\label{Fsf}
\end{eqnarray}
the Lagrangian (\ref{GRmnsf}) is quasi-invariant, since 
\begin{eqnarray}
\fl \delta_{\rho} L_{\rm SF} = d\left\{\frac{\kappa}{n-2} \rho^I \star 
(e_I\wedge e_J\wedge e_K)\wedge\left[ f\, R^{JK} [\omega]  + \frac{2 \Lambda}{(n-1)(n-2)} e^J \wedge e^K \right]\right. \nonumber \\
\left. +\frac{\kappa}{n-2} \rho^I (T_{IJ}+2\alpha V \eta_{IJ}) \star e^J  \right\}. \label{changeLsf}
\end{eqnarray}

Note that now the transformation of the frame $e^I$ has an extra term as compared with (\ref{gaugetrn}). However, this term vanishes for $f=\rm{constant}$. On the other side, the transformation of the connection is affected by the matter field in a significant manner. First, notice that the inverse of the function $f$ shows up in the cosmological term of~(\ref{Gsf}), which is the only difference with respect to its vacuum counterpart given by~(\ref{zetan}). Second and more importantly, now there is an extra term (\ref{Fsf}) coming from the presence of the scalar field that is composed of two factors, one of which depends on $f$, and the other involves the components of the energy-momentum tensor. Furthermore, observe that the transformation of the scalar field under the new gauge transformation is the same the one under a diffeomorphism generated by the vector field $\rho^I\partial_I$, showing that scalar fields are treated equally by both the new gauge symmetry and diffeomorphisms. To get the case of a scalar field minimally coupled to general
relativity we simply set $f=1$ and $K=1$, in whose case (\ref{Gsf}) is equal to (\ref{zetan}), while (\ref{Fsf}) keeps its dependency on the energy-momentum tensor. 

As we have mentioned, the Lagrangian (\ref{GRmnsf}) is invariant under local Lorentz transformations, for which the transformations of the frame and the connection are still given by (\ref{lorenztt}), whereas the scalar field remains invariant, that is, $\delta_{\tau} \phi = 0$. Note that the Noether identity corresponding to local Lorentz transformations is the same as~(\ref{loretzgaugeiden}), while the identity associated to (\ref{ngtsf}) is
\begin{eqnarray}
\fl (-1)^n D\mathcal{E}_I+\mathcal{E}_J\wedge\frac{1}{n-2}\partial_I\intprod [e^J\wedge d(\ln f)]+\mathcal{E}_{KL} \wedge (G_n{}^{KL}{}_{IJ}+F_n{}^{KL}{}_{IJ}) e^J+\mathcal{E}_{\phi}\partial_I\phi=0.\nonumber\\
\end{eqnarray}

\section{Conclusion}

By using the converse of Noether's second theorem, in this paper we have shown that the Palatini action for general relativity with or without cosmological constant in dimensions greater than three possesses a new internal gauge symmetry that is the higher-dimensional generalization of local translations in three dimensions. This result is something unexpected since diffeomorphisms are usually assumed to be the symmetry underlying general relativity, and, in fact, they are the symmetry that results after implementing Noether's procedure in the case of the Einstein-Hilbert action~\cite{Noether,EmmyNoether}. We have also reported the four-dimensional analog of this symmetry for the Holst action with cosmological term, showing that, off-shell, it depends on the Immirzi parameter [see (\ref{tresholstsym})]. Since infinitesimal diffeomorphisms can be written in terms of the new gauge symmetry and Lorentz transformations, in this framework diffeomorphisms are no longer considered fundamental and become a derived symmetry.  Thus, the full gauge invariance of general relativity can be equivalently described by the set of symmetries composed of the new gauge transformation and local Lorentz transformations, whose gauge algebra turns out to be open. As an application of our approach, we have also regarded the non-minimal coupling of a scalar field to gravity in $n$ dimensions, showing that this matter field affects the transformations of both the  frame and the connection.

Lastly, we have some comments:

\begin{enumerate}
	\item The action defined by the Holst term itself also possesses a gauge symmetry analogous to either (\ref{gaugetrn}) or (\ref{tresholstsym}), but now 
	\begin{eqnarray}
	\delta_{\rho}\omega^{IJ}=(R^{IJ}{}_{KL} + Y^{IJ}{}_{KL})\rho^Ke^L.
	\end{eqnarray}
	Here, as in the case of general relativity, the gauge algebra involving this symmetry and Lorentz transformations is open, whereas a similar relation exists among diffeomorphisms and them. 
	\item The gauge algebra of (\ref{ngtsf}) and the Lorentz transformation for the case of the non-minimal coupling of a scalar field to gravity  is currently being computed and will be reported elsewhere. Following this line of thought, it would be worth exploring the modifications induced on the new gauge symmetry by the coupling of Yang-Mills fields and fermions to general relativity. In this regard, we expect the matter fields to contribute to the transformation of the gravitational variables under the new gauge symmetry.
	\item Similarly, it would be very interesting to investigate the existence of the analog of the new internal gauge symmetry in other models of gravity. 
	\item The fact that the whole gauge symmetry of general relativity is purely internal renders it more analogous to an ordinary gauge theory, which might shed new insights into the quantization of gravity and open up new prospects to complete this program. On the one side, finding the canonical generator of the new symmetry could be useful for the canonical quantization program. On the other side, since the gauge algebra (\ref{algebrapalatnd1})-(\ref{algebrapalatnd3}) is open, we could try to implement the BV formalism~\cite{BATALIN198127,FUSTER-HENNEAUX2005}  for building the quantum theory associated to it. Furthermore, since the symmetry associated to the Holst action hinges on the Immirzi parameter, we expect it to play a significant role in the resulting quantum theory.
	
\end{enumerate}


\ack 
We thank Alejandro Perez, Carlo Rovelli, Jos\'e D. Vergara, and Jos\'e A. Zapata for their valuable comments. This work was supported in part by Consejo Nacional de Ciencia y Tecnolog\'{i}a (CONACyT), M\'{e}xico, Grant No. 237004-F.

\section*{References}
\bibliography{references}

\end{document}